\documentclass[aps,prd,preprint,nofootinbib]{revtex4}

\usepackage{graphicx}

\begin{document}

\title{An analysis of $H \to \gamma \gamma$ up to three-loop QCD corrections}

\author{Sheng-Quan Wang}
\author{Xing-Gang Wu}
\email{wuxg@cqu.edu.cn}
\author{Xu-Chang Zheng}
\author{Gu Chen}
\author{Jian-Ming Shen}

\address{Department of Physics, Chongqing University, Chongqing 401331, P.R. China}

\date{\today}

\begin{abstract}
The principle of maximum conformality (PMC) provides a convenient way for setting the optimal renormalization scales for high-energy processes, which can eliminate the conventional renormalization scale error via an order-by-order manner. At present, we make a detailed PMC analysis on the Higgs decay $H\rightarrow \gamma\gamma$ up to three-loop QCD corrections. As an important point of deriving reliable PMC estimation, it is noted that only those $\{\beta_i\}$-terms that rightly determine the running behavior of coupling constant via the renormalization group equation should be absorbed into the coupling constant, and those $\{\beta_i\}$-terms that pertain to the quark mass renormalization and etc. should be kept as a separate. To avoid confusion of separating and absorbing different types of $\{\beta_i\}$-terms into the coupling constant, we first transform the decay width in terms of top quark $\overline{\rm MS}$ mass into that of on-shell mass and then apply the PMC scale setting. After applying PMC scale setting, the final estimation is conformal and is scheme-independent and scale-independent. Up to three-loop QCD corrections, we obtain a PMC scale $\mu^{\rm PMC}_{r}=242.3$ GeV $\sim 2M_H$, which is optimal and highly independent of any choice of initial scale. Thus, we obtain a more accurate scale-independent prediction by taking the Higgs mass as the same as that of ATLAS and CMS measurements, i.e., $\Gamma(H\rightarrow \gamma\gamma)|_{\rm ATLAS}=9.504^{+0.226}_{-0.252}$ keV and $\Gamma(H\rightarrow \gamma\gamma)|_{\rm CMS}=9.568^{+0.195}_{-0.191}$ keV, where the error is caused by the measured Higgs mass, i.e. the Higgs mass $M_{H}$ is taken as $125.5\pm0.2^{+0.5}_{-0.6}$ GeV for ATLAS and $125.7\pm0.3\pm0.3$ GeV for CMS, respectively. \\

\noindent PACS number(s): 12.38.Bx, 14.80.Bn, 11.10.Gh


\end{abstract}

\maketitle

\section{Introduction} \label{sec:1}

The discovery of a new gauge boson has been reported by CMS and ATLAS collaborations at the Large Hadron Collider (LHC)~\cite{higgs1,higgs2,higgsmass1,sign2}. At present, its properties are found to be remarkably similar to Standard Model (SM) Higgs~\cite{sign1,sign2,cms131,atlas131,atlascms131,higgsmass1}. To know more of its properties becomes a paramount task, since if we know it well, we may decide whether there is really new physics beyond SM and to determine further what's the new physics could be like. The Higgs decay channel, $H\rightarrow \gamma\gamma$, which provides a dominant role for the discovery of Higgs boson, shall be of great importance for revealing detailed properties of Higgs boson at the LHC or future collider experiments. The recent updates on Higgs boson search show that the signal significance of $H\rightarrow \gamma\gamma$ channel are $3.2\sigma-3.9\sigma$ for CMS~\cite{cms131} and $6.1\sigma-7.4\sigma$ for ATLAS~\cite{atlas131}. The decay width $\Gamma_{H\to \gamma\gamma}$ also enters into the cross section for Higgs-boson production through photon-photon fusion, then, it could also be helpful for future high energy $e^+ e^-$ colliders as international linear collider (ILC)~\cite{ILC} or the Higgs factory. Therefore, it is important to study this decay channel as precise as possible.

Theoretically, within the SM model, the decay of Higgs boson into photons is mediated though either $W$ boson or heavy fermions at least at the one-loop level, whose total decay rate can be generally written as
\begin{eqnarray}
\Gamma(H\rightarrow \gamma\gamma)&=& \frac{M_{H}^{3}} {64\pi} \left|A_{W}(\tau_{W})+\sum\limits_{f} A_{f}(\tau_{f})\right|^{2}, \label{htolorr}
\end{eqnarray}
where $\tau_{W}=M^{2}_{H}/4M^{2}_{W}$ and $\tau_{f}=M^{2}_{H}/4m^{2}_{f}$ with $f=(t, b, c, \tau)$ corresponding to top quark, bottom quark, charm quark and $\tau$ lepton, respectively. The electroweak $W$ boson loop provides dominant contribution to the process, which is about $4.5$ times larger than that of the top-quark loop. The fermionic contributions are proportional to an overall factor $(m_f/M_H)^4$, which indicates that the fermionic contributions are dominated by top-quark loop and the contributions from the light fermions as $b$ or $c$ quarks can be safely neglected, especially for their higher loop corrections.

Because of its importance, in the literature, many efforts have been made on studying the Higgs decays into two photons, e.g. the one-loop estimation can be found in Refs.\cite{corlo1,corlo2}, the two-loop estimation can be found in Refs.\cite{corr1,corr2,corr3,corr4, corr5,corr6,corr7,corr8,corr9,corr10}, and the electroweak corrections at the two-loop level have also been analyzed in Refs.\cite{ele1,ele2,ele3,ele4}. Most importantly, three-loop non-singlet QCD correction has been given in Ref.\cite{nonsing}, and recently, a complete three-loop correction including both singlet and non-singlet QCD corrections has been done by Ref.\cite{nonandsi}. All those improvements provide us a great chance for deriving more accurate estimation on Higgs properties, as is the purposes of our present paper.

It is conventional to choose a typical momentum transfer of the process as the renormalization scale and take an arbitrary range to estimate the uncertainty in the QCD prediction. As is well-known, under such conventional scale setting, there is renormalization scale and renormalization scheme ambiguities at any finite order. It is often argued that by varying the renormalization scale,
one can estimate contributions from higher-order terms.
However, this procedure only exposes the $\{\beta_i\}$-dependent nonconformal
terms, not the entire perturbative series. And the prediction of conventional scale setting is usually wrong in QED where there is never a scale uncertainty. Further more, the value of $n_f$ entering the QCD $\beta$-function is not determined by using conventional scale setting. A recent review on this point can be found in Ref.\cite{pmc8}. It has been suggested that the principle of maximum conformality (PMC) provides a possible solution for eliminating renormalization scale ambiguity~\cite{pmc1,pmc2,pmc3,pmc4,pmc5,pmc9,pmc6}, which also provides a principle to set the optimal renormalization scales up to all orders and to set the value of $n_f$ entering the QCD $\beta$-function at each perturbative order. In the present paper, we shall adopt PMC to eliminate the renormalization scale ambiguity for $H\rightarrow \gamma\gamma$. For the purpose, one can first finish the renormalization procedure by using an arbitrary initial scale $\mu_r^{\rm init}$, and then set the effective or optimal PMC scales for the process by absorbing all non-conformal terms into the coupling constant via an order-by-order manner. It is noted that the PMC satisfies all self-consistency conditions for setting the renormalization scale, and the QCD predictions under PMC are then independent of renormalization scheme. Because of these merits, the PMC method can be widely applied to high energy physics processes, some of its applications have already been shown in the literature~\cite{pmc2,pmc3,pmc4,pmc9,pmc6,jpsi,pom,higbbgg}.

In principle, the scale dependence of the strong coupling constant is controlled by the renormalization group equation (RGE) via the $\beta$ function, i.e.,
\begin{equation} \label{basic-RG}
\beta(\alpha_s)=\frac{d}{d\ln\mu^2_r}\left(\frac{\alpha_s(\mu_r)}{4\pi}\right) =-\sum_{i=0}^{\infty}\beta_{i}\left(\frac{\alpha_s(\mu_r)}{4\pi}\right)^{i+2} ,
\end{equation}
where various terms in $\beta_0$, $\beta_1$, $\cdots$, correspond to one-loop and two-loop $\cdots$ contributions respectively. If one can find a proper way to sum up all known-type of $\{\beta_i\}$-terms into the coupling constant, then one can determine the effective coupling for a specific process definitely at each perturbative order, and thus, the renormalization scale dependence can be greatly suppressed or even be eliminated. The PMC has been designed for such purpose. As an important point of deriving reliable PMC estimation, it is noted that only those non-conformal $\{\beta_i\}$-terms that rightly determine the running behavior of the coupling constant via RGE should be absorbed into the coupling constant to form a conformal series. Those unrelated $\{\beta_i\}$-terms that are from the quark mass renormalization and etc. should be kept as a separate during PMC scale setting. Since all $\{\beta_i\}$-terms are entangled with each other, one should find a proper way to deal with each types via a confidential way. In this paper, by suggesting an unambiguous way to carefully deal with the $\{\beta_i\}$-terms, we shall make a detailed PMC analysis on Higgs decay $H\rightarrow \gamma\gamma$ up to three-loop QCD corrections.

The remaining parts of this paper are organized as follows. In Sec.\ref{sec:2}, we present the calculation technology for the Higgs decay $H\rightarrow \gamma\gamma$ up to three-loop QCD corrections under the PMC scale setting, in which we suggest a new way to set the PMC scales for high energy processes unambiguously. In Sec.\ref{sec:3}, we present the numerical results and discussions. The Sec.\ref{sec:4} is reserved for a summary.

\section{Calculation technology}
\label{sec:2}

The PMC scales of the process can be determined by absorbing $\{\beta_{i}\}$-terms into the coupling constant. It is noted that the three-loop expressions presented in Ref.\cite{nonandsi} are calculated under the $\overline{\rm MS}$ scheme and are given for the top quark $\overline{\rm MS}$ mass ($m_{t}$). Then, the $\{\beta_i\}$-terms for both the top-quark anomalous dimension and the QCD $\beta$-function are entangled with each other. This makes it hard to set PMC scales in an unambiguous way. To solve the problem, we suggest to transform expressions in the terms of $m_{t}$ into those of the top quark on-shell mass ($M_{t}$). After doing such transformation, all remaining $\{\beta_i\}$-terms are rightly pertained to the coupling constant and the PMC scales can be readily determined. Practically, this transformation can be achieved by using the relation between $m_{t}$ and $M_{t}$~\cite{mstopo,mb2,mb3}. More specifically, their relation up to ${\cal O}(\alpha_s^3)$ can be written as
\begin{eqnarray}
m_{t}(\mu_{r})&=&M_{t}\bigg[1+\bigg(-{4\over 3}-\ln{\mu^{2}_{r}\over M^{2}_{t}}\bigg){\alpha_{s}(\mu_{r})\over \pi} +{1\over 288} \bigg(-3161
 + 142 n_{f} - 112 \pi^2 \nonumber\\
&&\quad\quad +16 n_{f} \pi^2 - 32 \pi^2 \ln{2}
-1884  \ln{\mu^{2}_{r}\over M^{2}_{t}} +
   104 n_{f} \ln{\mu^{2}_{r}\over M^{2}_{t}}-252 \ln^{2}{\mu^{2}_{r}\over M^{2}_{t}}\nonumber\\
&&\quad\quad + 24  n_{f}\ln^{2}{\mu^{2}_{r}\over M^{2}_{t}} + 48 \zeta_{3}\bigg)\bigg({\alpha_{s}(\mu_{r})\over \pi}\bigg)^{2}+ {\cal O}\left(\left({\alpha_{s}(\mu_r) \over \pi}\right)^{3}\right)\bigg]. \label{msmasspo}
\end{eqnarray}

After applying such transformation from $m_t$ to $M_t$ on the decay width for $H\rightarrow \gamma\gamma$, we can rewrite the decay width into the following schematic form
\begin{equation}
\Gamma(H\rightarrow \gamma\gamma) = \frac{M_{H}^{3}} {64\pi}\left[A_{\rm LO} + A_{\rm NLO}(\mu^{\rm init}_{r}){\alpha_{s}(\mu^{\rm init}_{r})\over \pi} +A_{\rm NNLO}(\mu^{\rm init}_{r}) \left({\alpha_{s}(\mu^{\rm init}_{r})\over \pi}\right)^2 + A_{\rm EW} {\alpha\over \pi} \right] , \label{hrrconv}
\end{equation}
where $\mu^{\rm init}_r$ stands for an arbitrary initial choice of renormalization scale~\footnote{Before applying PMC scale setting, we should first transform the expressions with full initial scale dependence~\cite{pmc8}, whose value is generally arbitrary. The elimination of scale dependence is equivalent to the elimination of initial scale dependence.}, and under conventional scale setting, it is usually fixed to be typical momentum of the process, e.g. $\mu_r \equiv \mu^{\rm init}_r =m_H$. The coefficients, similar to Ref.\cite{nonandsi} are defined as,
\begin{eqnarray}
A_{\rm LO}&=&\bigg(A_{W}^{(0)}(\tau_{W})+\hat A_t A_{t}^{(0)}(\tau_{t})+A_{f}^{(0)}(\tau_{f})\bigg)^{2}, \\
A_{\rm NLO}(\mu^{\rm init}_{r})&=&2\sqrt{A_{\rm LO}}\hat A_t A_{t}^{(1)}(\tau_{t}), \\
A_{\rm NNLO}(\mu^{\rm init}_{r})&=&2\sqrt{A_{\rm LO}} \; Re\bigg[\hat A_t A_{t}^{(2)}(\tau_{t})\bigg]+\bigg(\hat A_t A_{t}^{(1)}(\tau_{t})\bigg)^{2},\\
A_{\rm EW}&=& 2\sqrt{A_{\rm LO}}A^{(1)}_{\rm EW} ,
\end{eqnarray}
where $\hat A_t = N_c\frac{ 2 \alpha \sqrt{\sqrt{2} G_F}}{3 \pi} Q_t^2$. The one-loop functions $A_{W}^{(0)}(\tau_{W})$ and $A_{f}^{(0)}(\tau_{f})$, together with $A^{(1)}_{\rm EW}$ denoting the electroweak corrections to the W boson and top-quark induced processes, are free from strong coupling and should be kept as constants when applying PMC, whose explicit forms are~\cite{corlo1,corlo2}
\begin{eqnarray}
A_{W}^{(0)}(\tau_{W})&=&-\frac{\alpha\sqrt{\sqrt{2}G_{F}}} {2\pi} \left [2+{3\over \tau_{W}}+{3\over \tau_{W}} \left(2-{1\over \tau_{W}}\right) f(\tau_{W})\right],\\
A_{f}^{(0)}(\tau_{f})&=& N_c\frac{\alpha \sqrt{\sqrt{2} G_F}}{\pi \tau_{f}} Q_f^2 \left[1+\left(1-{1\over \tau_{f}}\right) f(\tau_{f})\right], \label{htolorrat}
\end{eqnarray}
where $N_{c}=3$, $G_F$ is the Fermi constant, $Q_f$ denotes the quark electric charge and
\begin{eqnarray}
f(\tau)&=&\left\{
\begin{array}{ll}
\arcsin^2(\sqrt{\tau}) &\quad \mbox{for}\quad \tau\leq 1\\
-\frac{1}{4}\left[\ln\frac{1+\sqrt{1-\tau^{-1}}}{1-\sqrt{1-\tau^{-1}}}-i\pi\right]^2&
\quad \mbox{for}\quad \tau>1
\end{array}\right. \;.
\end{eqnarray}
The $A_{f}^{(0)}(\tau_{f})$ with $f=(c, b, \tau)$ corresponding to charm quark, bottom quark and $\tau$ loop contributions at LO. For the case of $\tau$, we should set $N_c=1$. The coefficients $A_{\rm NLO}(\mu^{\rm init}_{r})$ and $A_{\rm NNLO}(\mu^{\rm init}_{r})$ are scale dependent, in which $A_{t}^{(i)}(\tau_{t})$ with $i=(0, 1, 2)$ are from one-, two-, and three-loop QCD contributions. Since the value of $\tau_t\sim 0.1$, we can expand $A_{t}^{(i)}(\tau_{t})$ in power series of $\tau_t$. The one-loop $A_{t}^{(0)}(\tau_{t})$ and two-loop $A_{t}^{(1)}(\tau_{t})$ can be expanded in the following forms up to order ${\cal O}(\tau_{t}^{6})$,
\begin{eqnarray}
A_{t}^{(0)}(\tau_{t})&=&1+\tau_{t}\bigg({7\over 30}\bigg)+\tau_{t}^{2}\bigg({2\over21}\bigg)+\tau_{t}^{3}\bigg({26\over525}\bigg)
+\tau_{t}^{4}\bigg({512\over 17325}\bigg)+\tau_{t}^{5}\bigg({1216\over 63063}\bigg) , \\
A_{t}^{(1)}(\tau_{t})&=&-1+\tau_{t}\bigg({122\over135}\bigg)+\tau_{t}^{2}\bigg({8864\over 14175}\bigg)+\tau_{t}^{3}\bigg({209186\over 496125}\bigg)
+\tau_{t}^{4}\bigg({696616\over2338875}\bigg) +\tau_{t}^{5}\bigg({54072928796\over 245827456875}\bigg) .
\end{eqnarray}
The three-loop contribution $A_{t}^{(2)}(\tau_{t})$ includes two parts, one is the non-singlet part $A_{t,0}^{(2)}(\mu^{\rm init}_{r})|_{\rm ns}$ and the other is the singlet part, which can be further divided into $A^{(2)}_{t,t}|_{\rm sin}$ for the terms with two top-quark loops and $A^{(2)}_{t,q}|_{\rm sin}$ for the terms with one top-quark loop and one light-quark loop. That is,
\begin{eqnarray}
A_{t}^{(2)}(\tau_{t}) &=& A_{t,0}^{(2)}(\mu^{\rm init}_{r})|_{\rm ns} + A^{(2)}_{t,t}|_{\rm sin} + \left(Q^{-2}_{t}\sum \limits_{q\neq t}Q^{2}_{q} \right) A^{(2)}_{t,q}|_{\rm sin} ,
\end{eqnarray}
where $Q_{q}$ is the electromagnetic charge of the light quark $q\in\{u, d, s, c, b\}$. Because the singlet parts $A^{(2)}_{t,t}|_{\rm sin}$ and $A^{(2)}_{t,q}|_{\rm sin}$ do not have $\{\beta_{i}\}$-terms (or $n_{f}$-terms), so they are also kept as constant during the PMC scale setting. Their analytic expressions for $\mu_r \equiv\mu^{\rm init}_r$ can be derived from Ref.\cite{nonandsi}. Here, for brevity, we present $A^{(2)}_{t,t}|_{\rm sin}$ and $A^{(2)}_{t,q}|_{\rm sin}$ in numerical form by setting $M_{H}=126$ GeV and $M_{t}=172.64$ GeV, i.e.
\begin{eqnarray}
A^{(2)}_{t,t}|_{\rm sin} = 0.15121- 0.00305i, \;\;
A^{(2)}_{t,q}|_{\rm sin} = -0.35836+0.52883 i .
\end{eqnarray}
Up to order ${\cal O}(\tau_{t}^{6})$, the non-singlet part $A_{t,0}^{(2)}(\mu^{\rm init}_{r})|_{\rm ns}$ in the terms of on-shell mass $M_{t}$ can be written as
\begin{eqnarray}
A_{t,0}^{(2)}(\mu^{\rm init}_{r})|_{\rm ns}&=&A_{t,0}^{(2)}(\mu^{\rm init}_{r})|^{n_{f}}_{\rm ns}\cdot n_{f}+A_{t,0}^{(2)}(\mu^{\rm init}_{r})|^{\rm con}_{\rm ns} \nonumber\\
&=& n_{f} \bigg(-{1\over18} + {1\over6} \ln{(\mu^{\rm init}_{r})^{2}\over M_{t}^{2}}\bigg) -{23\over24} -{11\over 4} \ln{(\mu^{\rm init}_{r})^{2}\over M_{t}^{2}} +
\tau_{t}\bigg[ n_{f}\bigg(-{47041\over 124416} - {7 \pi^2\over 270}\nonumber\\
&& - {61\over405} \ln{(\mu^{\rm init}_{r})^{2}\over M_{t}^{2}} + { 2681 \zeta_{3}\over9216}\bigg) + \bigg(-{18120683\over 622080} + {49 \pi^2\over 270} + {7\pi^2\over 135} \ln2 \nonumber\\
&& +{671\over 270}\ln{(\mu^{\rm init}_{r})^{2}\over M_{t}^{2}} + {772805 \zeta_{3}\over 27648} \bigg) \bigg] +
\tau^{2}_{t} \bigg[ n_{f}\bigg(-{12504637\over34836480} - {4 \pi^2\over189}\nonumber\\
&& - {  4432\over42525}\ln{(\mu^{\rm init}_{r})^{2}\over M_{t}^{2}} + {61397 \zeta_{3}\over221184}\bigg) + \bigg(-{51082579973\over2612736000} + {4 \pi^2\over27} +{ 8 \pi^{2}\over189 } \ln2 \nonumber\\
&&+ {  24376 \over14175}\ln{(\mu^{\rm init}_{r})^{2}\over M_{t}^{2}} + {66912377 \zeta_{3}\over 3317760}\bigg) \bigg]+ \tau^{3}_{t} \bigg[  n_{f} \bigg(-0.3364 - {26 \pi^2\over 1575} \nonumber\\
&&-0.0703\ln{(\mu^{\rm init}_{r})^{2}\over M_{t}^{2}}+0.2852  \zeta_{3}\bigg)  +  \bigg(-827.988 + {26 \pi^2\over 225}+ { 52 \pi^2\over 1575}\ln2 \nonumber\\
&& +1.1595 \ln{(\mu^{\rm init}_{r})^{2}\over M_{t}^{2}} +691.894\zeta_{3}\bigg) \bigg]+ \tau^{4}_{t} \bigg[ n_{f} \bigg(-0.3133 - {2048 \pi^2\over 155925} -0.0496\ln{(\mu^{\rm init}_{r})^{2}\over M_{t}^{2}} \nonumber\\
&& +0.2897\zeta_{3}\bigg)  + \bigg(-2408.83 + {2048 \pi^2\over22275} + { 4096 \pi^2 \over 155925} \ln2+0.8191\ln{(\mu^{\rm init}_{r})^{2}\over M_{t}^{2}}\nonumber\\
&& +2006.25  \zeta_{3} \bigg) \bigg] + \tau^{5}_{t} \bigg[ n_{f}  \bigg(-0.2949 - {6080 \pi^2\over567567}-0.0367 \ln{(\mu^{\rm init}_{r})^{2}\over M_{t}^{2}} +0.2930\zeta_{3}\bigg) \nonumber\\
&&+ \bigg(-17838.5 + {  6080 \pi^2\over 81081} + {12160 \pi^2 \over 567567} \ln2+0.6049\ln{(\mu^{\rm init}_{r})^{2}\over M_{t}^{2}}+14841.7 \zeta_{3}\bigg) \bigg],
\end{eqnarray}
where $A_{t,0}^{(2)}(\mu^{\rm init}_{r})|^{n_{f}}_{\rm ns}$ and $A_{t,0}^{(2)}(\mu^{\rm init}_{r})|^{\rm con}_{\rm ns}$ stand for non-conformal and conformal terms of $A_{t,0}^{(2)}(\mu^{\rm init}_{r})|_{\rm ns}$, respectively.

With the help of the above expressions, we obtain the conformal and non-conformal terms for the three-loop coefficient $A_{\rm NNLO}(\mu^{\rm init}_{r})$, i.e.
\begin{eqnarray}
A_{\rm NNLO}(\mu^{\rm init}_{r})&=& A^{n_{f}}_{\rm NNLO}(\mu^{\rm init}_{r})\cdot n_{f}+A^{\rm con}_{\rm NNLO}(\mu^{\rm init}_{r}),
\end{eqnarray}
where the non-conformal $A^{n_{f}}_{\rm NNLO}(\mu^{\rm init}_{r})$ and the conformal $A^{\rm con}_{\rm NNLO}(\mu^{\rm init}_{r})$ can be written as
\begin{eqnarray}
A^{n_{f}}_{\rm NNLO}(\mu^{\rm init}_{r})&=&2\sqrt{A_{\rm LO}}\hat{A}_{t}A_{t,0}^{(2)}(\mu^{\rm init}_{r})|^{n_{f}}_{\rm ns}, \\
A^{\rm con}_{\rm NNLO}(\mu^{\rm init}_{r}) &=&2\sqrt{A_{\rm LO}} \; Re\bigg[\hat{A}_{t}\bigg(A_{t,0}^{(2)}(\mu^{\rm init}_{r})|^{\rm con}_{\rm ns}+A^{(2)}_{t,t}|_{\rm sin} +\bigg(Q^{-2}_{t}\sum \limits_{q\neq t}Q^{2}_{q} \bigg)A^{(2)}_{t,q}|_{\rm sin}\bigg)\bigg] +\bigg(\hat A_t A_{t}^{(1)}(\tau_{t})\bigg)^{2}.
\end{eqnarray}

After applying the standard PMC scale setting procedures to the decay width (\ref{hrrconv}), especially the $n_f$ terms are absorbed into the coupling constant with a combined form of $\beta_0\;(=11 - \frac{2}{3} n_f)$, the decay width can be simplified as,
\begin{widetext}
\begin{eqnarray}
\Gamma(H\rightarrow \gamma\gamma)&=& \frac{M_{H}^{3}} {64\pi}\bigg[A_{\rm LO} +{\alpha_{s}(\mu^{\rm PMC}_{r})\over \pi} A_{\rm NLO}(\mu^{\rm init}_{r}) \nonumber\\
&& \quad\quad\quad +\bigg({\alpha_{s}(\mu^{\rm PMC}_{r})\over \pi} \bigg)^2 \left({33\over 2} A^{n_{f}}_{\rm NNLO}(\mu^{\rm init}_{r})+A^{\rm con}_{\rm NNLO}(\mu^{\rm init}_{r})\right)+{\alpha\over \pi}A_{\rm EW}\bigg], \label{htorrpmc}
\end{eqnarray}
\end{widetext}
where the PMC scale
\begin{eqnarray}
\mu^{\rm PMC}_{r}&=& \mu^{\rm init}_{r} \exp\bigg[{3A^{n_{f}}_{\rm NNLO}(\mu^{\rm init}_{r})\over A_{\rm NLO}(\mu^{\rm init}_{r})} \bigg]. \label{htorrpmcscale}
\end{eqnarray}
it is noted that at the present three-loop level, we have only $\beta_0$-terms, so we have one PMC scale. It is noted that only the initial scale dependent logarithmic terms should be absorbed into the coupling constant simultaneously with the non-conformal terms via RGE, and the remaining conformal coefficients are still at the initial scale $\mu^{\rm init}_r$. More over, as will be shown later, the PMC scale $\mu^{\rm PMC}_{r}$ only formally depends on the choice of the initial renormalization scale, its own value and hence the decay width are almost independent of the initial choice of renormalization scale, and then the renormalization scale dependence can be greatly suppressed or even eliminated at the three-loop level.

\section{Numerical results and discussions}
\label{sec:3}

To do numerical calculation, we take Higgs mass $M_{H}=126$ GeV as its central value, the W boson mass $M_{W}=80.385$ GeV and $\tau$ lepton mass $M_{\tau}=1.78$ GeV, the Fermi constant $G_{F}=1.16637\times10^{-5}\rm GeV^{-2}$, and $\alpha=1/137$. We adopt two-loop $\alpha_{s}$ running with the fixed point $\alpha(M_{Z})=0.1184$~\cite{pdg} to set its corresponding $\Lambda_{\rm QCD}$, i.e., we obtain $\Lambda^{(n_f=3)}_{\rm QCD}=0.386$ GeV, $\Lambda^{(n_f=4)}_{\rm QCD}=0.332$ GeV, $\Lambda^{(n_f=5)}_{\rm QCD}=0.231$ GeV and $\Lambda^{(n_f=6)}_{\rm QCD}=0.0938$ GeV. By adopting the on-shell quark masses $M_{t}=172.64$ GeV, $M_{b}=4.78$ GeV and $M_{c}=1.67$ GeV~\cite{pdg,mb1,mb2}, we obtain the $\overline{\rm MS}$-running masses $m_{t}(M_{H})=166.43$ GeV, $m_{b}(M_{H})=2.73$ GeV and $m_{c}(M_{H})=0.62$ GeV.

\begin{table}[htb]
\centering
\begin{tabular}{|c|c|c|c|c|c|}
\hline
& \multicolumn{5}{c|}{Conventional scale setting} \\
\hline
~~~~& ~~$i$=LO~~ & ~~$i$=NLO~~ & ~~$i$=NNLO~~ & ~~$i$=QED~~& ~~$i$=Total~~\\
\hline
~~$\Gamma_i \;(10^{-3}\;\rm keV)$~~& ~~ 9650.3 ~~ & ~~ 162.0 ~~ & ~~ 2.2  ~~ & ~~$-148.0$~~ & ~~ 9666.5 ~~\\
\hline
& \multicolumn{5}{c|}{PMC scale setting}\\
\hline
~~~~& ~~$i$=LO~~ & ~~$i$=NLO~~ & ~~$i$=NNLO~~ & ~~ $i$=QED ~~ & ~~$i$=Total~~\\
\hline
~~$\Gamma_i \;(10^{-3}\;\rm keV)$~~& ~~ 9650.3~~ & ~~148.7~~&~~14.1 & ~~$-148.0$~~ &~~9665.2 \\
\hline
\end{tabular}
\caption{Decay width for the decay $H\rightarrow\gamma\gamma$ up to three-loop level, in which $\mu^{\rm init}_{r}=M_{H}$. For the conventional scale setting, the renormalization scale is fixed to be $\mu^{\rm init}_{r}$. $\Gamma_{\rm LO}$, $\Gamma_{\rm NLO}$, $\Gamma_{\rm NNLO}$ and $\Gamma_{\rm QED}$ denote the decay widths at LO, NLO, NNLO levels and QED correction, respectively. $\Gamma_{\rm Total}$ stands for the total decay width up to NNLO level. } \label{tababpmc}
\end{table}

We present the decay width for $H\rightarrow \gamma\gamma$ before and after PMC scale setting in Table \ref{tababpmc}, where $\Gamma_{\rm LO}$, $\Gamma_{\rm NLO}$ and $\Gamma_{\rm NNLO}$ denote the decay widths at LO, NLO and NNLO levels accordingly, and $\Gamma_{\rm Total}$ stands for the total decay width up to NNLO level. We adopt $\mu^{\rm init}_{r}=M_{H}$. Under conventional scale setting, the renormalization scale $\mu_r \equiv \mu^{\rm init}_{r}$, while after PMC scale setting, the renormalization scale is determined by Eq.(\ref{htorrpmcscale}). Total decay width remains almost unchanged before and after PMC scale setting, which is because about $98\%$ contribution comes from the LO terms that are free from strong interactions. After the PMC scale setting, the NLO contribution changes down from $1.65\%$ to $1.52\%$, which indicates that the resummation of $\beta_0$-terms up to all orders shall give negative contributions to NLO terms. From the Table \ref{tababpmc}, one may observe that the ratio $K=\Gamma_{\rm NNLO}/\Gamma_{\rm NLO}$ amounts to be only about $1\%$ of the NLO correction, which changes to $9\%$ for PMC. This, however, does not mean that the pQCD convergence of the conventional scale setting is better than that of PMC \footnote{After applying PMC, the convergent renormalon terms are resummed into the coupling constant and the pQCD convergence shall be improved in principle. With more types of renormalon terms, or more loop corrections, being included, such improvement will be more clear. }. In fact, because of large scale uncertainties for both the NLO and NNLO decay widths under the conventional scale setting, the value of $K$ shall be varied within the region of $[-9\%,+10\%]$ even for a small variation $\mu^{\rm init}_{r}\in[M_H/2,2M_H]$, where the minus sign means the NNLO terms are negative.

\begin{table}[htb]
\centering
\begin{tabular}{|c|c|c|c|}
\hline
~~ ~~ & \multicolumn{3}{c|}{$\Gamma_{\rm NLO}$ ($10^{-3}$ keV)} \\
\hline
~~~$\mu^{\rm init}_{r}$~~~ & ~~~$M_{H}/2$~~~ & ~~~$M_{H}$~~~ & ~~~$2 M_{H}$~~~ \\
\hline
Conventional scale setting & 180.1 & 162.0 & 148.0 \\
\hline
PMC scale setting & 148.7 & 148.7 & 148.7\\
\hline
\end{tabular}
\caption{Scale dependence for the NLO decay width $\Gamma_{\rm NLO}$, where three choices of $\mu^{\rm init}_{r}$ are adopted. It shows that the decay width under conventional scale setting shows a strong scale dependence. After the PMC scale setting, it is almost independent of $\mu^{\rm init}_{r}$. }
\label{scaleun}
\end{table}

When applying the PMC scale setting, after resumming all $\beta_0$-terms into the coupling constant, the results will be much more steady over the scale changes. As a comparison, the scale dependence for the NLO decay width $\Gamma_{\rm NLO}$ are presented in Table \ref{scaleun}, where three scales $\mu^{\rm init}_{r}=M_{H}/2$, $M_{H}$, $2M_{H}$ are adopted. The decay width of the NLO QCD corrections under the conventional scale setting shows a strong dependence on $\mu^{\rm init}_{r}$, i.e. the scale errors for the NLO correction terms are $[+11\%,-8.6\%]$ for $\mu^{\rm init}_{r}\in[M_H/2,2M_H]$. Furthermore, under the conventional scale setting, the NNLO decay width also shows a much larger scale uncertainty, i.e. $\Gamma_{\rm NNLO}=\left(2.2^{-19}_{+12}\right) \times 10^{-3}$ keV for $\mu_r\in[M_H/2,2M_H]$. In contrast, under the PMC scale setting, the decay width of NLO QCD corrections is almost unchanged. This is because that the PMC scale itself is fixed and highly independent of $\mu^{\rm init}_{r}$.

\begin{figure}[htb]
\centering
\includegraphics[width=0.48\textwidth]{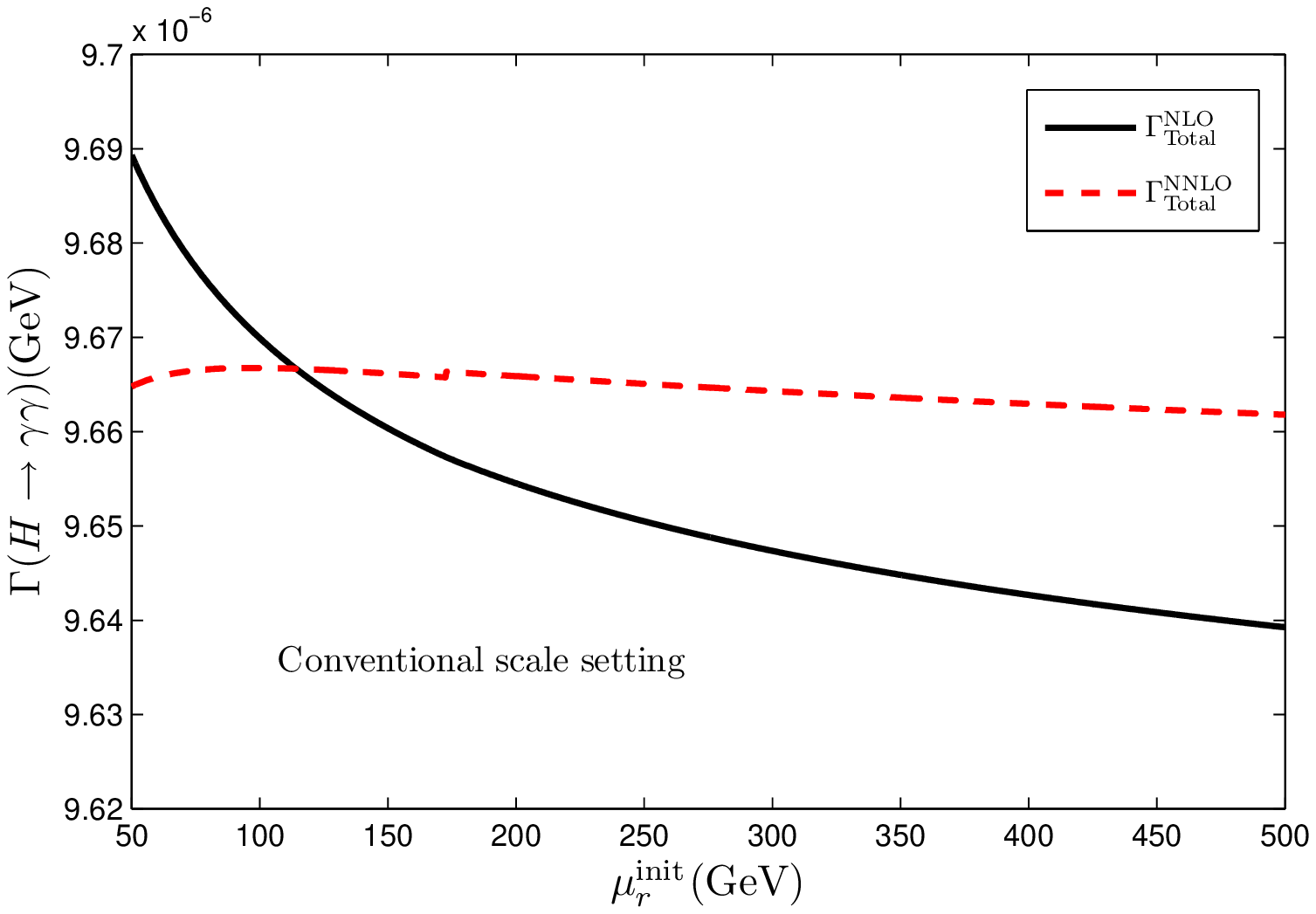}
\includegraphics[width=0.48\textwidth]{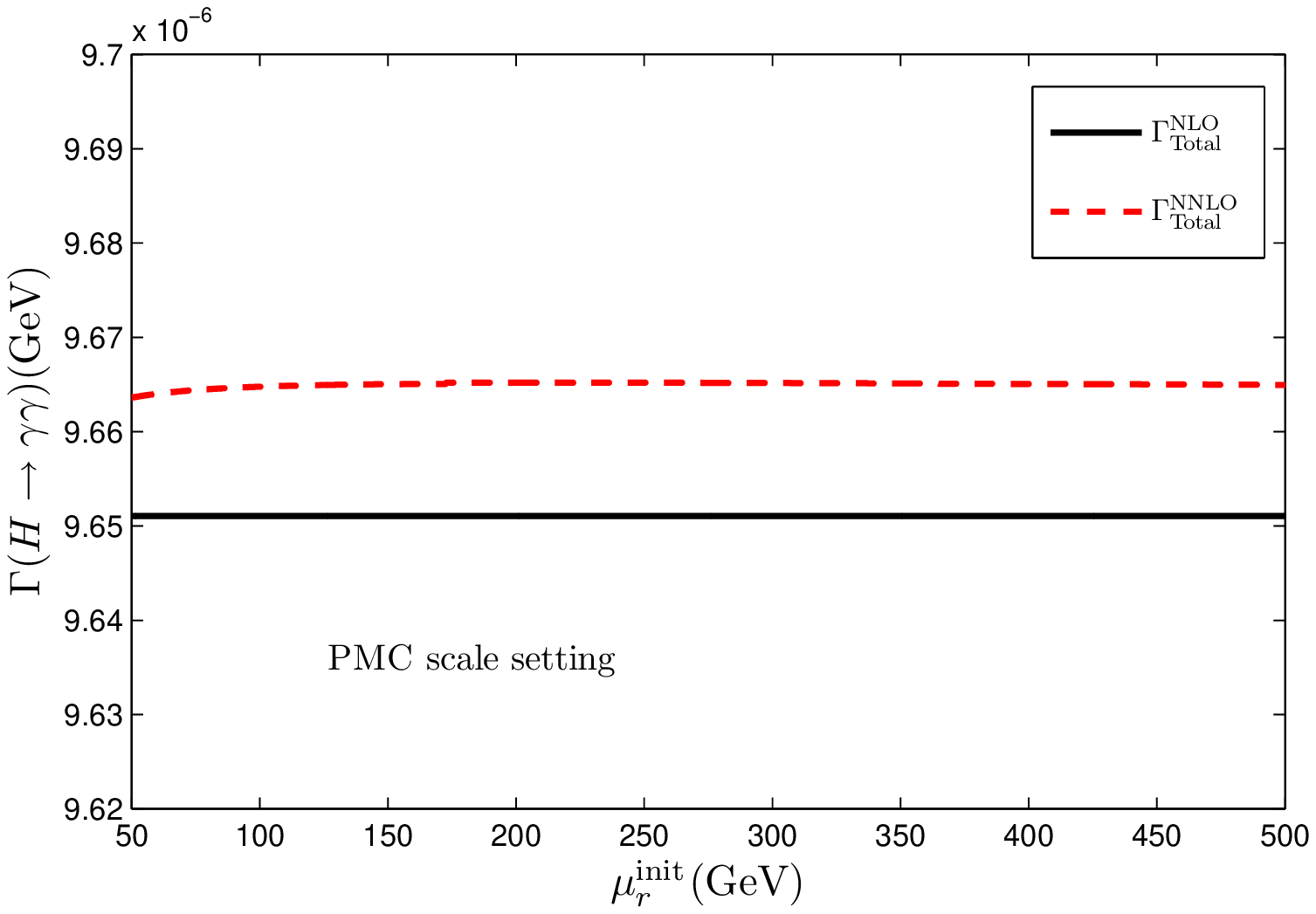}
\caption{Total decay width versus the initial renormalization scale $\mu^{\rm init}_{r}$ for $H\rightarrow\gamma\gamma$ under conventional scale setting and PMC scale setting. The solid and dashed lines stand for the total decay widths up to NLO level and NNLO level, respectively.} \label{Plot:sqrst}
\end{figure}

More explicitly, by using Eq.(\ref{htorrpmcscale}), we find that the PMC scale is fixed to be $242.3$ GeV $(\sim 2 M_H)$ for any choice of $\mu^{\rm init}_{r}$. This shows that the PMC scale is larger than the typical momentum flow of the process, i.e. the Higgs mass $M_{H}$, to a certain degree. This larger scale is consistent with the above observations that after PMC scale setting the NLO correction to the decay width becomes smaller than that of the conventional scale setting. This point can be shown explicitly in Fig.(\ref{Plot:sqrst}), in which the solid and dashed lines stand for the total decay widths up to NLO level ($\Gamma_{\rm Total}^{\rm NLO}$) and NNLO level ($\Gamma_{\rm Total}^{\rm NNLO}$), respectively. The PMC estimation as shown by the right diagram of Fig.(\ref{Plot:sqrst}) shows that the decay widths at each perturbative order are almost flat versus $\mu^{\rm init}_{r}$. In principle, at different pQCD orders, we need to set different PMC scales~\cite{pmc8}. At present, the NNNLO calculation is not available, then we have no $\{\beta_{i}\}$-terms to determine the PMC scale for NNLO terms, so we set the PMC scale for NNLO terms to be equal to the PMC scale for NLO terms, i.e. $\mu^{\rm PMC}_{r}|_{\rm NNLO}=\mu^{\rm PMC}_{r}|_{\rm NLO}=\mu^{\rm PMC}_{r}$, which is calculated by Eq.(\ref{htorrpmcscale}). This treatment will cause residual scale dependence after PMC scale setting~\cite{pom}. To provide a relatively reliable estimation on such residual scale dependence, following the idea of PMC, we first rewrite the NNLO coupling constant in Eq.(\ref{htorrpmc}) as follow
\begin{displaymath}
{\alpha_{s}(\mu^{\rm PMC}_{r})\over \pi} = {\alpha_{s}(\mu^{\rm init}_{r})\over \pi}+{\beta_{0}\over 4} \ln\left({\mu^{\rm init}_{r}\over \mu^{\rm PMC}_{r}}\right)^{2}\left({\alpha_{s}(\mu^{\rm init}_{r}) \over \pi}\right)^{2} .
\end{displaymath}
We observe that such newly log-term at one-order higher, corresponding to part of the NNNLO level, can compensate the scale changes at the NNLO level and result in an improved scale error in comparison to the conventional way~\cite{jpsi}.

Thus, after PMC scale setting, we can eliminate the renormalization scale ambiguity and obtain a more accurate predications in comparison to the previous estimations for the Higgs decay $H\rightarrow\gamma\gamma$~\cite{hrr1,hrr2}. As a comparison, we present the decay widths at each perturbative order under the conventional scale setting with pole top-quark mass and under the PMC scale-setting in the following. That is, by using Eqs.(\ref{hrrconv},\ref{htorrpmc}), we have
\begin{eqnarray}
\Gamma(H\rightarrow \gamma\gamma)|_{\rm pole}&=& \left[9.502+ \left(1.620^{+0.180}_{-0.140}\right) \times10^{-1}+ \left(2.200^{-18.585}_{+12.486}\right) \times10^{-3}\right]\; {\rm keV}, \label{hrrmspo}\\
\Gamma(H\rightarrow \gamma\gamma)|_{\rm PMC}&=& \left[9.502 +1.487\times10^{-1}+1.415\times10^{-2}\right]\; {\rm keV}, \label{hrrmspmc}
\end{eqnarray}
where the errors in Eq.(\ref{hrrmspo}) is caused by varying initial scale $\mu^{\rm init}_r$ from $M_{H}/2$ to $2M_{H}$. Under the conventional scale setting, the decay width of the NNLO terms increases with the increment of the scale, while the decay width of the NLO terms decreases with the increment of the scale, then as a combination, the total decay width up to the NNLO level varies slightly with the change of scale as shown in Fig.(\ref{Plot:sqrst}). After PMC scale setting, the residual scale dependence for the NNLO terms is less than $0.1\%$, so our PMC estimation (\ref{hrrmspmc}) shows no scale dependence even at the present three-loop order.

As a final remark. For a pQCD estimation, it is helpful to predict what's the ``unknown" QCD corrections could be and how it will be changed with the improved QCD corrections. The conventional estimation done by varying the scale over a certain range is not proper, since it can only estimate the non-conformal contribution but not the conformal one. More over, after the PMC scale setting, the determined PMC scales are optimal and can not be varied simply, otherwise, it will explicitly break the renormalization group invariance and lead to unreliable estimation. To achieve an estimation of how the `'unknown" QCD corrections could be from the ``known" QCD corrections, we rewrite the decay width as
\begin{equation}
\Gamma(H\rightarrow \gamma\gamma) = \frac{M_{H}^{3}} {64\pi}\left[A_{\rm LO} \left(1 + \tilde{R}_n \right) + A_{\rm EW} {\alpha\over \pi} \right] . \label{hrrconvnew}
\end{equation}
Up to $n$-loop QCD correction, $\tilde{R}_n=\sum_{i=1}^{n}\tilde{\cal C}_i  a^{i}_s$, where $a_s=\alpha_s/\pi$ and the known coefficients $\tilde{\cal C}_i$ for the convention and PMC scale settings can be read from Eqs.(\ref{hrrconv},\ref{htorrpmc}). For example, under the conventional scale setting, $\tilde{\cal C}_1=\frac{A_{\rm NLO}(\mu^{\rm init}_{r})}{A_{\rm LO}}$ and $\tilde{\cal C}_2=\frac{A_{\rm NNLO}(\mu^{\rm init}_{r})}{A_{\rm LO}}$. We suggest the following more conservative method for the scale error analysis or to estimation how the ``unknown" higher-order QCD corrections could be; i.e. to take the scale uncertainty as the last known perturbative order. More explicitly, the pQCD uncertainty at the $(n+1)$-order is $\Delta_n=\pm |{\tilde{\cal C}}_{n} a^{n}_s|_{\rm MAX}$, where both ${\tilde{\cal C}}_{n}$ and $a_s$ are calculated by varying the initial scale to be within the region of $[M_H/2,2M_H]$ and the symbol ``MAX'' stands for the maximum value of $|{\tilde{\cal C}}_{n} a^{n}_s|$ within this region. This treatment is natural for PMC, since after PMC scale setting, the pQCD convergence is ensured and the only uncertainty is from the last term due to the unfixed PMC scale at this particular order. We obtain $\Delta_1\sim\pm1.9\times10^{-2}$ and $\Delta_2\sim\pm1.7\times10^{-3}$ for the conventional scale setting; $\Delta_1\sim\pm1.4\times10^{-2}$ and $\Delta_2\sim\pm1.4\times10^{-3}$ for PMC. The error bars provide a consistent estimate of the ``unknown" QCD corrections under various scale settings; i.e., the exact value for the ``unknown" $\tilde{R}_{2}$ are well within the error bars predicted from $\tilde{R}_{1}$. A detailed discussion on how to estimate the ``unknown" QCD corrections for various scale settings from $R(e+e-)$ and $H\to b\bar{b}$ up to four-loop level shall be presented elsewhere~\cite{new}.

\section{Summary}
\label{sec:4}

We have applied the PMC scale setting to study the decay width of $H\rightarrow\gamma\gamma$ up to three-loop level. After the PMC scale setting, we obtain a renormalization scale and renormalization scheme independent estimation. Then, a more accurate estimation has been achieved. More explicitly,

\begin{itemize}
\item The PMC renormalization scale is formed by absorbing the $\{\beta_{i}\}$-terms that govern the running behavior of the coupling constant into the coupling constant. At present, the decay width has been calculated up to three-loop level by using the $\overline{\rm MS}$ renormalization scheme. Because the $\{\beta^{\overline{\rm MS}}_i\}$-terms for both the top-quark anomalous dimension and the QCD $\beta$-function are entangled with each other, it is hard to apply PMC unambiguously without knowing each part well. Thus, to separate the $\{\beta_{i}\}$-terms in an unambiguous and convenient way, we transform the expressions with the $\overline{\rm MS}$-mass $m_t$ to those with the pole-mass $M_{t}$ before applying the PMC scale setting. It is noted that, our present suggestion of dealing with the $\{\beta_i\}$-terms can also be extended to deal with other source terms involving $\{\beta_i\}$-series that are also unrelated to determine the running behavior of the coupling constant in order to derive a reliable PMC estimation.

\item As shown by the left diagram of FIG.(\ref{Plot:sqrst}), under the conventional scale setting, the NLO decay width $\Gamma_{\rm NLO}$ shows a strong dependence on $\mu^{\rm init}_{r}$, one can obtain a convergent and less scale dependent estimations up to three-loop level. Especially, there is large scale cancelation between different perturbative terms, i.e. the decay width of the NNLO terms increases while the decay width of the NLO terms decreases with the increment of the scale. In contrast, as shown by the right diagram of FIG.(\ref{Plot:sqrst}), after the PMC scale setting, the decay widths at the each perturbative order are almost flat versus the initial renormalization scale $\mu^{\rm init}_{r}$. It shows that the optimal renormalization scale for $H\to\gamma\gamma$ is $\sim 2M_H$ other than the usual adopted $M_H$. It is noted that because the NLO level PMC scale can be definitely determined by the $\{\beta_0\}$-terms at the NNLO level, the total decay width up to NLO level is almost flat as shown by Table \ref{scaleun}; while there is no $\{\beta_i\}$-terms to set the PMC scale for NNLO terms, so there is residual scale dependence for NNLO terms, which is quite small via an improved conventional scale setting method as suggested in Ref.\cite{jpsi}.

\begin{figure}[htb]
\centering
\includegraphics[width=0.48\textwidth]{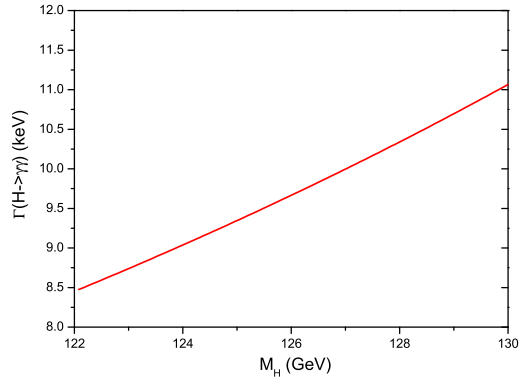}
\caption{Decay width of $H\rightarrow\gamma\gamma$ versus the Higgs mass $M_{H}$ after the PMC scale setting. } \label{Plot:sqpms:c}
\end{figure}

\item Since the scale dependence has been eliminated, after the PMC scale setting, a more accurate scale-independent pQCD predication for $H\rightarrow\gamma\gamma$ can be obtained. By taking the region for the Higgs mass as the same as the ones determined by ATLAS and CMS collaborations, the decay width under the PMC scale setting are as follows,
    \begin{eqnarray}
     \Gamma(H\rightarrow \gamma\gamma)|_{\rm ATLAS} &=& 9.504^{+0.226}_{-0.252} \; {\rm keV}, \\
     \Gamma(H\rightarrow \gamma\gamma)|_{\rm CMS} &=& 9.568^{+0.195}_{-0.191} \; {\rm keV},
    \end{eqnarray}
    where the subscript ${\rm ATLAS}$ means the error is caused by varying Higgs mass $M_{H}=125.5\pm0.2^{+0.5}_{-0.6}$ GeV determined by the ATLAS collaboration and the subscript ${\rm CMS}$ means the error is caused by varying Higgs mass $M_{H}=125.7\pm0.3\pm0.3$ GeV determined by the CMS collaboration, respectively. More explicitly, we show the decay width versus the Higgs mass $M_{H}$ in FIG.(\ref{Plot:sqpms:c}). In addition to previous examples done in the literature, the PMC scale setting works well for the decay channel $H\rightarrow\gamma\gamma$.  Inversely, if taking the decay width as an input, one can obtain a more accurate estimation on the Higgs mass. Thus, it is helpful to reveal the properties of Higgs boson with high precision, and it can also increase our understanding of possible new physics beyond the SM.
\end{itemize}

It is noted that a four-loop estimation of $H\to\gamma\gamma$ has been presented very recently~\cite{new1}. It, being given with top-quark $\overline{\rm MS}$-mass, can not be directly dealt with by PMC. However, we can estimate that the four-loop terms shall not affect our main conclusions since our present three-loop estimation already shows good pQCD convergence and it is reasonable to estimate that the four-loop terms shall provide small contribution after applying PMC scale setting.

\hspace{1cm}

{\bf\Large Acknowledgments}: We thank Stanley J. Brodsky and Matin Mojaza for helpful discussions. This work was supported in part by Natural Science Foundation of China under Grant No.11275280, by the Program for New Century Excellent Talents in University under Grant No.NCET-10-0882, and by the Fundamental Research Funds for the Central Universities under Grant No.CQDXWL-2012-Z002.

\end{document}